\documentclass[aip,cha,amsmath,amssymb,superscriptaddress,
preprint,%
author-year,%
cha]{revtex4-2}
\usepackage{graphicx}
\usepackage{dcolumn}
\usepackage{bm}

\usepackage[utf8]{inputenc}
\usepackage[T1]{fontenc}
\usepackage{mathptmx}
\usepackage{etoolbox}
\usepackage{subfigure}

\makeatletter
\def\@email#1#2{%
 \endgroup
 \patchcmd{\titleblock@produce}
  {\frontmatter@RRAPformat}
  {\frontmatter@RRAPformat{\produce@RRAP{*#1\href{mailto:#2}{#2}}}\frontmatter@RRAPformat}
  {}{}
}%

\makeatother
\usepackage[version=3]{mhchem}
\usepackage{amsthm,url}

\usepackage{amsmath,amsthm,bm}
\usepackage{enumerate}
\usepackage{hyperref}
\usepackage{svg}
\usepackage{rotating}
\usepackage{xcolor,comment}

\theoremstyle{definition}
\newtheorem{lemma}{Lemma}
\newtheorem{proposition}{Proposition}
\newtheorem{definition}{Definition}

\newcommand{\Pf}[1]{{\textit{Proof.}\ }#1}

\newcommand*{\emp}[1]{\textbf{#1}}

\newcommand*{\alphab}{\bm{\alpha}}
\newcommand*{\betab}{\bm{\beta}}
\newcommand*{\gammab}{\bm{\gamma}}

\newcommand*{\cb}{\bm{c}}
\newcommand*{\kb}{\bm{k}}

\newcommand*{\rateb}{\mathbf{rate}}

\newcommand*{\rth}{r^{\mathrm{th}}}
\newcommand*{\T}{^{\top}}

\newcommand*{\janos}[1]{{\color{red}#1}}
\newcommand*{\matyi}[1]{{\color{blue}#1}}

\begin{document}

\preprint{AIP/123-QED}

\title[Chemical chaos]{Rigorously proven chaos in chemical kinetics}
\thanks{The authors extend their deepest gratitude to Prof. V. G\'asp\'ar for his ongoing assistance, constructive criticism, and encouragement. 
The present work has been supported by the National Research, Development, 
and Innovation Office for funding (FK-134332).}

\author{M. Susits}
\email{susits.matyas@gmail.com}
\affiliation{ 
Department of Analysis and Operations Research, Institute of Mathematics, Budapest University of Technology and Economics, M\H{u}egyetem rkp. 3., H-1111 Budapest, Hungary.}
\author{J. T\'oth}%
\email{jtoth@math.bme.hu}
\affiliation{ 
Department of Analysis and Operations Research, Institute of Mathematics, Budapest University of Technology and Economics, M\H{u}egyetem rkp. 3., H-1111 Budapest, Hungary}.
\affiliation{Chemical Kinetics Laboratory, Institute of Chemistry, E\"otv\"os Lor\'and University, P\'azm\'any P\'eter sét\'any 1/A, H-1117 Budapest, Hungary.}
\date{\today}

\begin{abstract}
This study addresses a longstanding question regarding the mathematical proof of chaotic behavior in kinetic differential equations. Following the numerous numerical and experimental results in the past 50 years, we introduce two formal chemical reactions that rigorously demonstrate this behavior. Our approach involves transforming chaotic equations into kinetic differential equations and subsequently realizing these equations through formal chemical reactions. The findings present a novel perspective on chaotic dynamics within chemical kinetics, thereby resolving a longstanding open problem.
\end{abstract}
\maketitle
\begin{quotation}

In the first half of the 20th century, researchers discovered that certain complex reactions, before reaching the stationary state, can exhibit oscillatory behavior representing a peculiar form of exotic dynamics.

Open systems not only have the capacity for sustained oscillations but also multistationarity. While formal reaction kinetics has yielded relevant general mathematical results regarding multistationarity and oscillation, the same cannot be said for chaotic behavior: there is a lack of comprehensive mathematical results in this area. However, both experiments and numerical simulations suggest that chaos may manifest in mass-action-type kinetic models. As of now, the exploration of chaotic behavior in this context remains largely reliant on experimental and numerical findings.
In our study, we have constructed formal reactions that exhibit chaos, and we have verified this chaotic behavior through symbolic analysis.
\end{quotation}

\section{Introduction}
In chemistry the prevailing belief historically and in contemporary times has been that chemical reactions follow a simple course. 
It was commonly held that once you initiate a reaction by introducing certain chemical species, 
the system tends (sooner or later monotonically) to a stable stationary state, 
called an equilibrium state in chemistry.
In this state, the concentrations of all involved species remain constant. 
These stationary concentrations can often depend on the initial concentrations of the reacting species, although some may be independent of these initial conditions (as seen in \cite{shinarfeinberg}).

However, in the early decades of the 20th century (see e.g. \cite{sharmanoyes}) and later in the 1950s (\cite{zhabotinsky}), researchers discovered that certain complex reactions---before being close to the stationary state---can exhibit (damped) oscillatory behavior, a peculiar form of what we might call \emp{exotic} behavior in the world of chemistry. 
Open systems can also display sustained oscillations or even multistationarity, meaning they can reach multiple distinct steady states. 
The concentrations in these stationary states can vary in intricate ways, 
sometimes forming intriguing shapes like mushrooms or isolas in models, see \cite{lili}, or even in experiments \cite{ganapathisubramanianshowalter}. 

Shortly after meteorologist Lorenz's groundbreaking observations (\cite{lorenz}), the question arose: can chaotic behavior manifest itself in the context of chemical reactions? 
This question, raised first by \cite{showalternoyesbareli} elicited two general responses. 

For chemists inclined toward theory, there is a consensus that chemical reactions can indeed exhibit all the hallmarks of chaos (\cite{scott}). 
Observations of sensitivity to initial conditions, period doubling, and unimodal Poincaré maps have been documented in various reactions. 
While there exists relevant general mathematchical results about
multistationarity, e.g. \cite{balakotaiahluss,joshishiu,voitiukpantea},
and oscillation, e.g. \cite{pota} and \cite{banajiboros},
no such work seems to exist about chaotic behavior. See Table \ref{tab:results}.
Even in the recent special issue of Physica D \cite{katsanikasagaoglou} titled "Chaos Indicators, Phase Space and Chemical Reaction Dynamics" there are no papers on chemical kinetics and the question of whether kinetic differential equations can show chaotic behavior is not even mentioned.
It is here that our present paper aims to make a convincing case.
We assert the statement -- which is an old conjecture from the mathematical sense -- 
that such models do exist and that the fundamental principles of standard mass action kinetics do not inherently preclude chaos.

 \begin{table}
     \centering
\begin{tabular}{|l|c|c|}
 \hline
    \textbf{Property} & \textbf{Theory} & \textbf{Experiment} \\
    \hline
    Multistationarity & x & x\\
    \hline
     Oscillation & x & x \\
     \hline
     Chaos & - & x \\
     \hline
 \end{tabular}
 \caption{\label{tab:results}Results on exotic phenomena in chemical kinetics.}
 \end{table}

Our construction strategy unfolds as follows. 
We begin with a system of polynomial differential equations known to exhibit key chaotic characteristics. 
Knowing a bound of a trapping region we shift the trajectories into the first orthant to avoid negative concentrations. 
Subsequently, we multiply the right-hand sides of these shifted equations by the product of some of the variables, resulting in a form akin to kinetic differential equations \cite{harstoth}, also known as \emp{Hungarian} equations,
i.e. polynomial differential equations without \emp{negative cross-effect}. 
Finally, we construct a realization of this equation in terms of reaction steps.

This is \emp{almost} the same as the method applied by \cite{poland}.
He started from the Lorenz equation
\begin{equation}\label{eq:lorenzorig}
\dot{x}=\sigma(y-x),\quad\dot{y}=\varrho x\boxed{-x z}-y,\quad\dot{z}=xy-b z
\end{equation}
with the boxed term expressing the fact that \(y\) is decreasing in a process in which it does not take part. 
Then, he shifted the coordinate functions to remain in the first orthant.
Next, he applied a transformation proposed by \cite{samardzijagrelleswasserman} leaving the time behavior of the coordinate functions qualitatively the same, or the trajectories exactly the same. Finally, he constructed a complex chemical reaction inducing the given differential equation. 

Why do we not accept this as the final solution to the problem of finding a formal chemical reaction with an induced kinetic differential equation showing chaotic behavior?
\begin{enumerate}
\item 
The quantity of shifts has been determined \emp{by visual observation of figures}, and it has not been proven that the values chosen will certainly shift the coordinate functions into the first orthant.
\item
To realize the term \(-x z\) expressing negative cross-effect the author introduced six ("fast" and "slow") reaction steps. 
This extension---with appropriate values of the parameters---will \emp{approximately} give the corresponding term.
\end{enumerate}

We start with the fact that the models by
Lorenz \cite{galiaszgliczynski,mischaikowmrozek1995,mischaikowmrozek1998},
Chen (\cite{zhoutangchen,zhengchen}),
L\u (\cite{luchen} 
have been rigorously proven to be chaotic, although we shall only deal in detail with the Lorenz and the Chen model as there is no estimation of the attractors available in the other cases.
In the case of the Lorenz model, the existence of a Smale horseshoe has been rigorously proven which implies
chaos according to the Def. \ref{def:devaney} of Devaney in the Appendix.
As for the Chen model, it has been proven \cite{zhoutangchen} that the Shil'nikov criterion holds which implies the existence of a Smale horseshoe.

The critical steps we focus on are as follows: 
\begin{enumerate}
\item 
\emp{exact} determination of the shifts needed to move the trajectories into the open first orthant, 
\item 
transforming the shifted equations into a kinetic (or Hungarian) type equation, and finally, 
\item 
finding a realization of the obtained kinetic differential equation in terms of reaction steps endowed with mass action type kinetics. 
\end{enumerate}

The structure of our paper is as follows. 
Subsection \ref{sec:chemical} reviews different characteristics of chaos found in real chemical reactions or models coming from experimental experience. It also shows a few formal kinetic models that seemingly show chaotic behavior, approximately. 
To arrive at our main results, in Section \ref{sec:rigor} we also follow a modified---rigorous---version of Poland's scenario to construct a formal chemical realization of the Lorenz model \eqref{eq:lorenzorig}. 
In a less detailed way, we shortly follow the same construction
for another, similar model by Chen.
Finally, Section \ref{sec:disc} summarizes the results and formulates open problems to be solved.
The two subsections of the Appendix summarize the basic concepts of formal reaction kinetics and
several definitions of chaos, without being a tutorial.
\section{Chaos in chemical kinetic experiments and models}\label{sec:chemical}
\cite{olsendegn} were the first to show chaos experimentally in an enzyme reaction.  

To demonstrate the possibility of chaos, one must undertake the study of real chemical reactions and substantiate (using the inductive approach typical in natural sciences) several criteria. These include confirming that concentration-versus-time curves exhibit aperiodicity and sensitivity to minor initial concentration perturbations, ensuring trajectories remain within a bounded set. Additionally, the absence of sharp peaks in the Fourier transform of concentrations, an approximately two-dimensional attractor, the ability to construct a one-dimensional unimodal Poincaré map, a positive Lyapunov exponent indicating trajectory divergence (Definition \ref{def:physicslyapunov}), and the manifestation of period doubling (Definition \ref{def:perioddoubling}) are essential markers.

Let us note that in real chemical kinetics, the boundedness of the trajectories automatically follows because the law of atomic balance implies mass conservation. 
This is just the opposite of the case of formal chemical kinetics allowing in- and outflow, where it is a serious problem to prove that the trajectories remain in a compact set \cite{deaktothvizvari}.

In the following table we cite some of the papers with the presence of chaotic behavior in experiments.
Purely theoretical works are also mentioned in the Table \ref{tab:chemical}.

\begin{table}
\centering
\begin{tabular}{|l|l|l|l|l|}
\hline
Reference&System&Signs of chaos&Verification method&Remarks\\
\hline\cite{olsendegn}                &enzyme reaction&aperiodicity&visual observation  
    &experiment  \\ 
    &peroxidase&&of data &\\
\hline\cite{schmitzgrazianihudson}&BZ reaction&aperiodicity&visual observation &experiment\\
&&Poincaré map&of data &\\
&&strange attractor&&\\
\hline\cite{willamowskirossler}&extended&strange attractor&numerics&formal kinetic model\\
& Lotka-Volterra&&&irreversible steps\\
&&&&not mass-conserving\\
\hline\cite{gyorgyituranyifield}&BZ reaction&strange attractor&numerics&experiment-motivated\\
&chaotic Oregonator&period doubling&&model\\
&&Poincaré map&&not mass action\\
\hline\cite{pengscottshowalter}&three-variable&period doubling&numerics&formal kinetic\\
&autocatalator&Poincaré map&&model \\
&&strange attractor&&irreversible steps\\
\hline\cite{rabaiorban}&pH oscillation&aperiodicity&numerics&experiment-motivated\\
&&&&model\\
\hline\cite{rabaiexperimental}&pH oscillation&aperiodicity&visual observation&experiment\\
&&period doubling&of data&\\
&&attractors&&\\
\hline\cite{rabaimodel}&pH oscillation&aperiodicity&numerics&experiment-motivated\\
\cite{gaspartoth}&&strange attractor&&model\\
\hline
\cite{wangxiao} & 4 dimensional & homoclinic  & numerics & formal kinetic model \\
& Lotka--Volterra& bifurcation &&\\
\hline
\end{tabular}
\caption{\label{tab:chemical}Chaos in kinetic experiments and models}
\end{table}

Let us mention in passing that some non-mass action type models were also offered, e.g. the model by \cite{hudsonrossler}.
\cite{ferreiraferreiralinoporto} analyzed a non-mass action type model of glycolysis using Fourier transform of the concentrations written in Mathematica, thereby supporting aperiodicity. As opposed to non-mass action kinetics we do not mention non-kinetic models, even if they are of the polynomial form. 
R\"ossler produced a series of such models \cite{rosslercont,rosslersimplereaction,rosslerprototype,hudsonrosslerkillory} beyond the kinetic model \cite{willamowskirossler}. 
Part of the early works by R\"ossler has been shortly summarized by \cite{gaspard}.

{Beyond Rössler and Poland, who were interested in the possibility of chaos in kinetic differential equations, 
there is another line of research connected to our present work, which is the area of Lotka--Volterra systems. 
As it is known \cite[Subsection 6.4.1]{tothnagypapp}, the differential equations of these systems are kinetic differential equations.

Most researchers would not doubt the presence of chaos in the four-dimensional Lotka--Volterra system
\cite[Eq. (3.1)]{wangxiao} reading that paper and looking at its figures. However, the authors use 
numerical calculations instead of pure mathematical arguments at critical steps. 
There is no problem with the calculation of the equilibria (Eq. (3.2)),
or with the condition of Hopf bifurcation (Eq. (3.3)). 
However, the existence of the homoclinic orbit is shown using numerical simulations (not a computer assited proof). Hence the conditions of the Shil'nikov theorem are just shown numerically, not proven.

The reader should not misunderstand our arguments: the exemplary work by \cite{wangxiao}
shows how to use calculations and figures to give inductive arguments 
to support the presence of chaos in a kinetic differential equation, 
but it is not a mathematical proof.

\section{Rigorously proven chaos in formal kinetic models}\label{sec:rigor}
We are trying to construct chemical reactions where the chaotic behavior is proven. 

We shall use two steps: first, we shift the variables to have only positive values. 
Next, using the modification by Cr\u{a}ciun of the idea of Samardzija we multiply the right-hand sides with the product of some of the concentrations, to get rid of the negative cross-effects and at the same time not modify the trajectories, as opposed to only using the method by \cite{samardzijagrelleswasserman}.

To employ our method effectively, it is essential to work with chaotic polynomial systems that possess well-defined trapping regions. Over the last five decades, there has been a wealth of systems where chaos has been rigorously established. However, pinpointing the exact location of the chaotic attractor presents a distinct and formidable challenge. While numerical simulations often give the impression that a system will remain within a certain region, identifying the precise coordinates of the chaotic attractor and providing trapping regions around them is a complex problem.

For instance, consider the Lorenz system published in 1963 \cite{lorenz}, which required nearly 25 years to achieve this task \cite{leonov}.
Subsequently, numerous refinements were made to delineate the trapping region of the chaotic attractor, highlighting the considerable time and effort invested in this process, see \cite{leonovbuninkoksch} and the references therein.
The trapping region was constructed using Lyapunov functions, and the attractor's existence was proved by \cite{tucker} in 1999. 
Similarly, demonstrating the global boundedness of the Chen system was no small feat \cite{barbozachen}, and the global boundedness of the L\u system has only yielded partial results, without encompassing the original parameter set selected by L\u, even though it is much like the Lorenz and Chen systems \cite{zhangliaozhang}. That lack of an exact estimate of the attractor location prevented us from using the same procedure for the L\u system which we have done for the other systems.

To sum it up, the rigorous validation of trapping regions for chaotic attractors is a challenging undertaking, necessitating diverse approaches tailored to the characteristics of each individual system.

Our starting point is the Lorenz model. 
\subsection{The Lorenz reaction}\label{subsec:lorenz}
It has been shown both using direct calculations (\cite{tothhars}) and the theory of \emp{algebraic invariants} (\cite{halmschlagerszenthetoth}) developed by \cite{sibirsky} and his coworkers that the Lorenz equation cannot be transformed via orthogonal transformations into a kinetic differential equation; a small, although symbolic result. 

Here we aimed at positive results: 
we construct reactions with rigorously proven chaotic behavior following and improving the plan proposed by \cite{poland}. 
First, we need a trapping region for the chaotic attractor (finding lower bounds is enough, but results are usually for the boundedness of the attractors of chaotic systems). Then given these lower bounds we shall shift the system such that it will remain in the positive orthant if started from the shifted trapping region.

Once it is ensured that the trajectories will never reach 0 in any coordinate we can use the method proposed by \cite{samardzijagrelleswasserman} to multiply the equations' right-hand side by $x y z$. $\dot{u} = f(u)*s(u)$ where $s: \mathbb{R}^3 \longrightarrow \mathbb{R}^+$ is a continuously differentiable scalar function. This does not change the trajectories of the autonomous system, it is only a time transform.

Because of the Hungarian lemma \cite[Theorems 3.1--2]{harstoth}, the absence of negative cross-effect implies that the equation can be realized by complex chemical reactions,
meaning that the differential equation will be the induced kinetic differential equation of the reaction, assuming mass-action type kinetics.

Let us start with the \emp{Lorenz equation}
\eqref{eq:lorenzorig}
where the most often used values of the parameters are
\(\sigma = 10, \varrho = 28, \beta = 8/3.\)
\subsubsection{Trapping region of the Lorenz system}
The function 
\begin{equation*}
V(x,y,z):= \varrho x^2 + \sigma y^2 + \sigma(z - 2\rho)^2
\end{equation*}
is a Lyapunov function for the Lorenz system outside the ellipsoid \[D = \{(x,y,z)| (\varrho x^2 + y^2 + b(z - \varrho)^2 \leq \varrho^2 b) + \varepsilon\}\]
since
\begin{align*}
\dot{V} & = 2 \sigma \varrho x (y - x) + 2 \sigma y (\varrho x - y - xz) + 2\sigma(z-2\varrho)(xy - bz) \\
    & = 2\sigma [-\varrho x^2 - y^2 - bz^2 + 2b\varrho z] \\
    & = -2\sigma(\varrho x^2 + y^2 + b(z - \varrho)^2 - \varrho^2 b) \leq -2\sigma\varepsilon & \text{ if } (x,y,z) \notin D
\end{align*}

We can use this to get a trapping region by choosing a $c$ such that $D \in E =\{(x,y,z)| V(x,y,z) < c\}$. All trajectories must pass inwards through the boundary of $E$ since $\dot{V}$ is negative on the boundary. 

The bounds of the ellipsoid E in each direction are:
\begin{align*}
|x| \leq \sqrt{c/\varrho},\\
|y| \leq \sqrt{c/\sigma},\\
|z-2\varrho| \leq \sqrt{c/\sigma}
\end{align*}

Thus, a shift greater than $(\sqrt{\varrho b}, \varrho \sqrt{b}, 0)$ will ensure that the trajectories do not leave the positive quadrant.
In out concrete case with $\varrho = 28$, $\sigma = 10$, $b = 8/3$ choosing $c = 200^2$ is large enough \ref{fig:EcontainsD}. A shift of $(200/\sqrt{28}, 200/\sqrt{10}, 200/\sqrt{10}-56) = (37.7964, 63.2456, 7.2456)$ is enough.

If the shift of $x$ is at least as much as the shift of $y$ then the new equation can be of order 4 instead of 5; closer to chemical reality. We will use the shift $(100, 100, 10)$ to derive the chemical reaction system of order 4 and allow initial concentrations in the 
$E: 28(x-100)^2 + (y-100)^2 + \frac{8}{3}(z-10)^2 = \frac{6272}{3}$  ellipsoid.

\begin{figure}\label{fig:EcontainsD}
    \centering
    \includegraphics[width=0.7\linewidth]{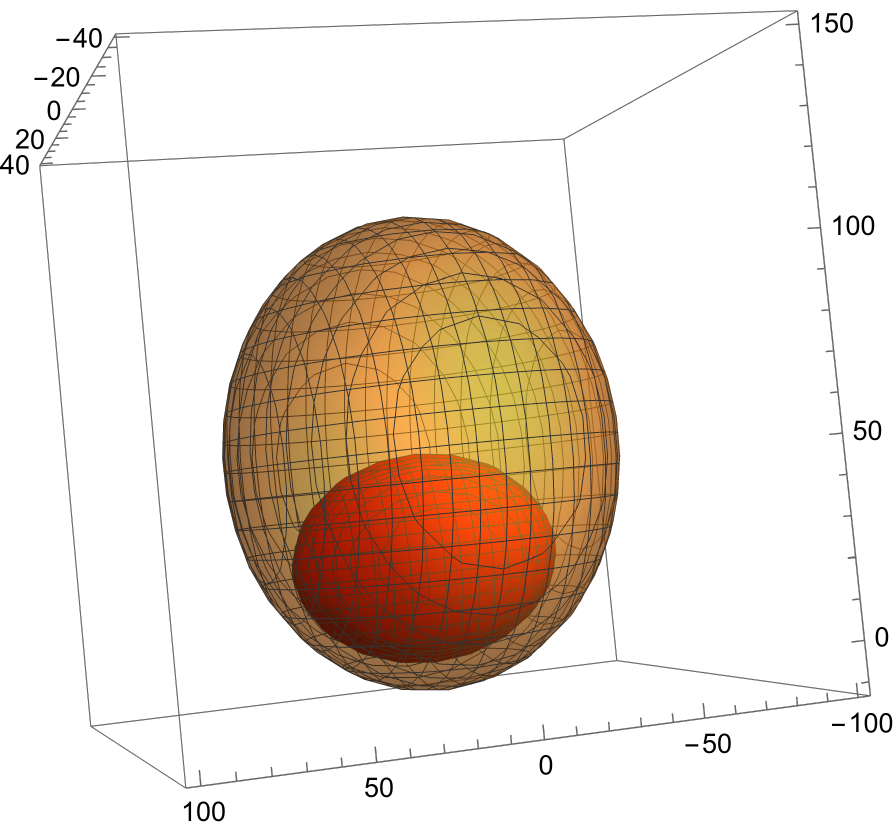}
    \caption{$D: 28x^2 + y^2 + \frac{8}{3}(z-28)^2 = \frac{6272}{3}$ and $E : 28x^2 + 10 y^2 + 10(z - 56)^2 = 200^2$}
    \label{fig:enter-label}
\end{figure}

We shall work with the initial conditions:
\(x(0) = 101, y(0) = 102, z(0) = 13.\)
As an illustration, Fig. \ref{fig:Lorenz} shows a component of the solution and also the trajectory in the 3D space foreshadowing the aperiodicity of the solution coordinates and the presence of a strange attractor.
We emphasize that all figures are used for visualization and illustration, not needed for all the mathematical proofs above.

\begin{figure}[!ht]
\centering
\includegraphics[width=0.45\linewidth]{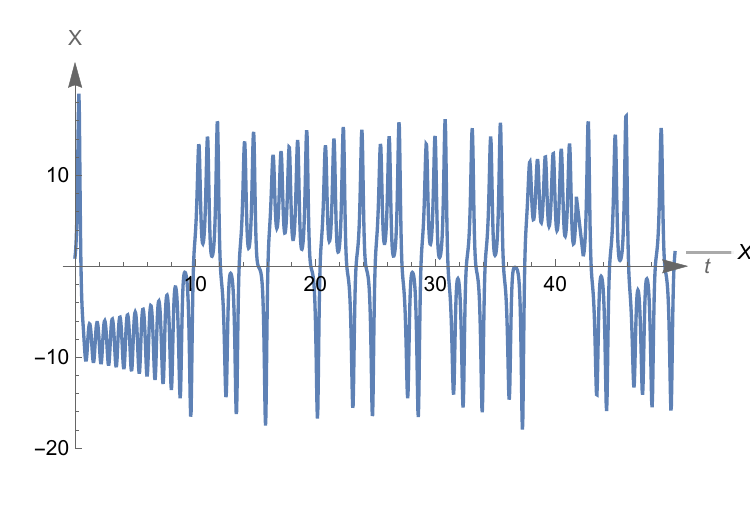}\quad
\includegraphics[width=0.45\linewidth]{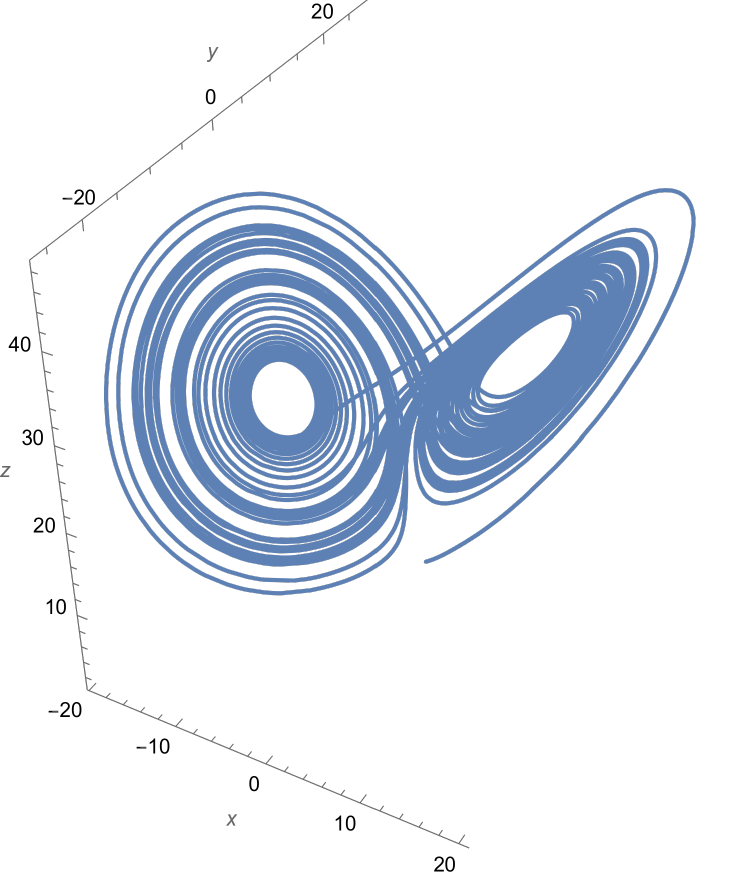}
\caption{Time evolution of the first component in the Lorenz equation with the parameters as given in the text (left) and the corresponding trajectory (right) started from $(1,2,3)$ on the time interval 0-50.}
\label{fig:Lorenz} 
\end{figure}
Leonov and his coworkers \cite{leonovbuninkoksch,zhangzhang} published a series of papers on the symbolic estimation of the location of the attractor. 
Here we only need a simple estimation: the attractor is surely contained in the sphere
with a radius of 31.76 centered at \((0,0,16.82)\). 
Therefore the trajectory will remain in the first orthant if one adds 
10, 50, and 0 to the coordinates, respectively, because they are confined
to the corresponding rectangular box.
Let us remark that for us no precise bounds are needed, however, researchers starting with Leonov and coworkers \cite{leonovbuninkoksch} provided a series of better and better estimates, finally one that is as good as the numerical bounds obtained originally by Lorenz.

\subsubsection{Transforming the Lorenz system}
The equation after this shift looks like this in the new variables \(X:=x+100, Y:=y+100, Z:=z+10\):
\begin{align}
X' &= \sigma(Y-X)\label{eq:LorenzShifted1} \\
Y' &= \varrho (X-100) -(X-100)Z = \varrho X \boxed{- XZ} + 100Z \boxed{- 100\varrho}\label{eq:LorenzShifted2} \\
Z' &= (X-100)(Y-100) -\beta Z = XY \boxed{-100X-100Y} - \beta Z + 10000\label{eq:LorenzShifted3}
\end{align}
There is no need to show figures now because they are simply obtained from \ref{fig:Lorenz} Fig. via shifting.
Eqs. \ref{eq:LorenzShifted2} and \ref{eq:LorenzShifted3} are still not a kinetic differential equation because they contain several terms expressing negative cross-effect, the boxed parts. We could multiply by $X Y Z$ from the second step, however in the case of this shifted Lorenz system multiplying by $Y Z$ is enough, making the system simpler by giving us reaction steps of order at most 4 instead of 5. 

The equations after substituting $\sigma = 10$, $\varrho = 28$, $\beta = 8/3$ and multiplying by $Y Z$ are the following 
\begin{align}
X' &= 10Y^2 Z - 10 XYZ,\label{eq:LorenzMultiplied01} \\
Y' &= 38XYZ - XYZ^2 -Y^2Z + 32YZ^2 - 3700YZ\label{eq:LorenzMultiplied02}\\
Z' &= XY^2Z - 100XYZ - 100Y^2 Z - 8/3 YZ^2 + (30080/3)YZ.\label{eq:LorenzMultiplied03}
 \end{align}
 
As a result of this transformation, one arrives at a system having the same trajectories in the first orthant and being kinetic at the same time. 
Because of the Hungarian lemma \cite[Theorems 3.1--2]{harstoth}, the absence of negative cross-effect implies that the equation can be realized by complex chemical reactions,
meaning that the differential equation will be the induced kinetic differential equation of a reaction, assuming mass-action type kinetics.
The realizations in both cases will be ugly in the sense that they contain reaction steps of a high order, some of them are not mass-conserving, they are not reversible, not even weakly reversible, let alone detailed balanced.  
\subsubsection{A realization of the transformed Lorenz system}
This is the (canonic, see \cite{harstoth}) realization of the transformed Lorenz equation \eqref{eq:LorenzMultiplied01}--\eqref{eq:LorenzMultiplied03}.

\begin{align}
&\ce{2 Y + Z ->[10] X + 2 Y + Z},\quad\ce{X + Y + Z ->[10] Y + Z},\label{eq:lorenz01}\\
&\ce{X + Y + Z ->[38] X + 2 Y + Z},\quad\ce{Y + 2Z ->[100] 2 Y + 2 Z},\label{eq:lorenz02}\\
&\ce{X + Y + 2 Z ->[1] X + 2 Z},\quad\ce{2 Y + Z ->[1] Y + 2 Z},\label{eq:lorenz03}\\
&\ce{Y + Z ->[3700] Z},\quad\ce{X + 2 Y + Z ->[1] X + 2 Y + 2 Z},\label{eq:lorenz04}\\
&\ce{Y + Z ->[30080/3] Y + 2 Z},\quad\ce{Z + Y + Z ->[100] X + Y},\label{eq:lorenz05}\\
&\ce{2 Y + Z ->[100] 2 Y},\quad\ce{Y + 2 Z ->[8/3] Y + Z},\label{eq:lorenz06}
\end{align}
Note that the way from kinetic differential equations to realizations by a reaction is not unique. 
The canonic representation can be obtained algorithmically, but in some cases, it is too complicated.

As an illustration, Fig. \ref{fig:LorenzMultiplied} shows a component of the solution and also the trajectory in the 3D space.
\begin{figure}[!ht]
\centering
\includegraphics[width=0.45\linewidth]{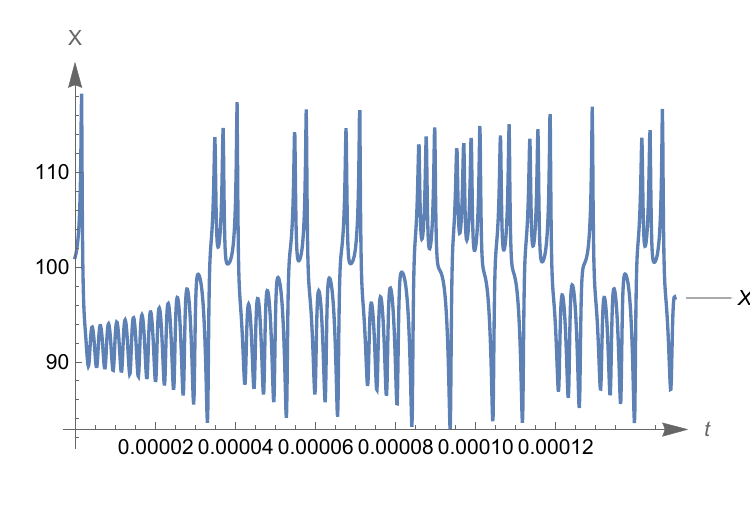}\quad
\includegraphics[width=0.45\linewidth]{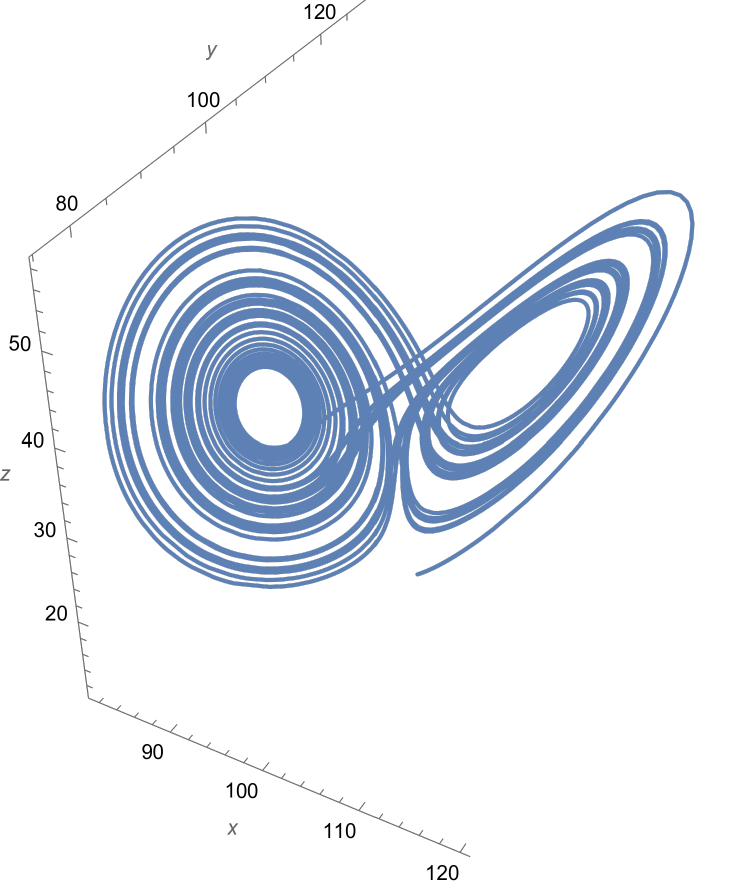}
\caption{Time evolution of the first component in the transformed Lorenz reaction (left) and the corresponding trajectory (right) started from $(101,102,13)$ on the time interval 0-0.00012.}
\label{fig:LorenzMultiplied} 
\end{figure}
Qualitative similarities (that have been proven just now!) are illustrated in the figures. 
Finally, one can have a look at the Feinberg--Horn--Jackson graph of the reaction \eqref{eq:lorenz01}--\eqref{eq:lorenz06}.
\begin{figure}[!ht]
\centering
\includegraphics[width=0.95\linewidth]{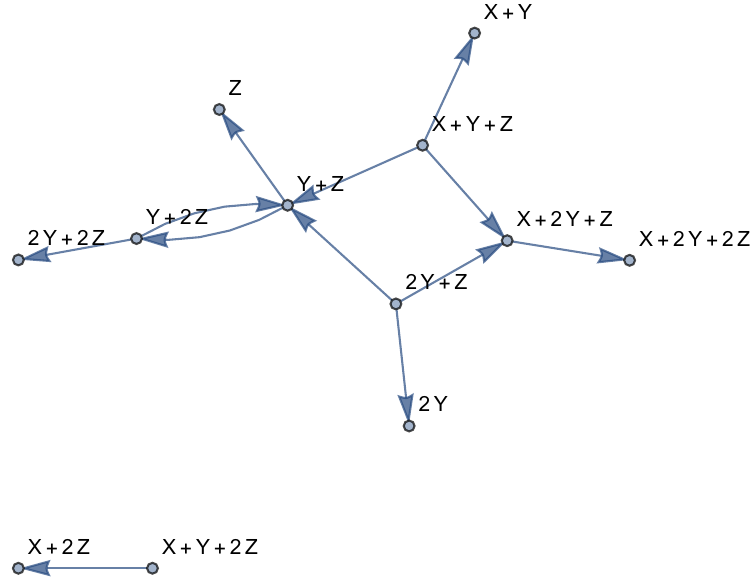}
\caption{The Feinberg--Horn--Jackson graph of the reaction \eqref{eq:lorenz01}--\eqref{eq:lorenz06}.
}
\label{fig:LorenzFHJ} 
\end{figure}
\subsection{The Chen reaction}\label{subsec:chen}
\cite{chenueta} proposed the differential equation with negative cross-effect:
\begin{equation}\label{eq:Chen}
\dot{x}=\sigma(y-x),\quad
\dot{y}=(c-a)x+cy\boxed{-x z},\quad
\dot{z}=xy-b z
\end{equation}
with the boxed term expressing the fact that \(y\) is decreasing in a process in which it does not take part. The usual values of the parameters are: \(\sigma=35, b=3, c=28.\) (A more general usual assumption is \(\sigma/2<c<\sigma.\))
The paper contains a single figure of the attractor and no proof at all. Later,
\cite{zhoutangchen,zhengchen} proved the existence of a Smale horseshoe.

\begin{figure}[!ht]
\centering
\includegraphics[width=0.45\linewidth]{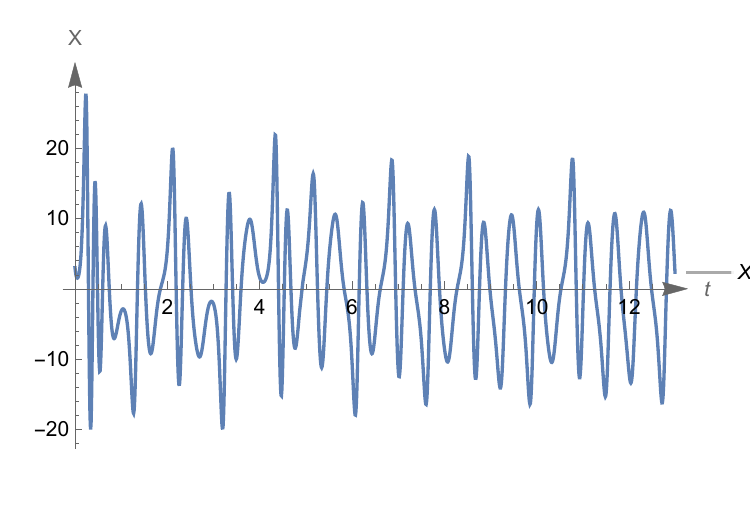}\quad
\includegraphics[width=0.45\linewidth]{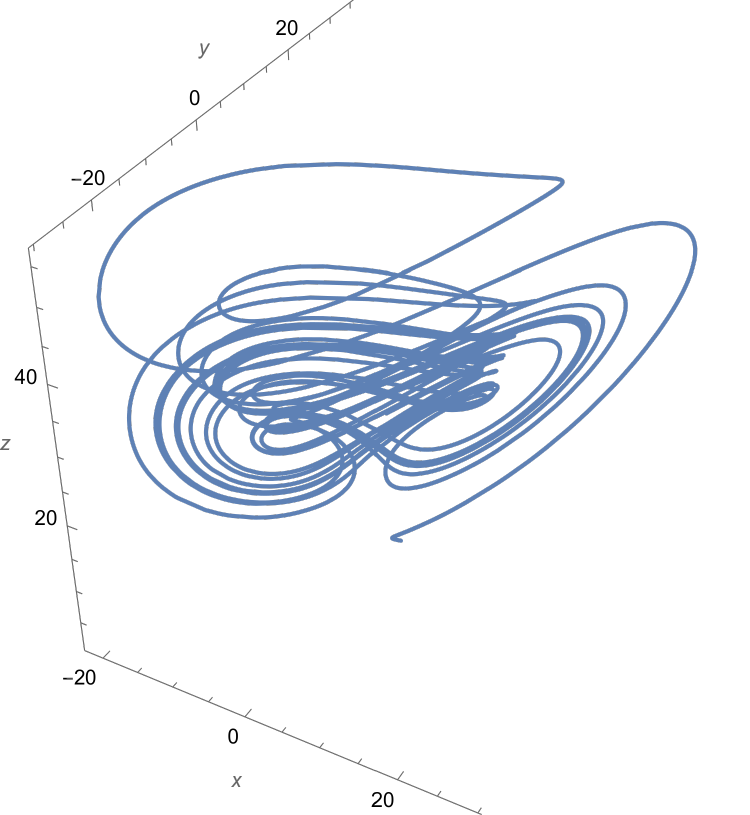}
\caption{Time evolution of the first component in the Chen equation with the parameters as given in the text (left) and the corresponding trajectory (right) started from $(3,1,4)$ on the time interval 0-13.}
\label{fig:Chen} 
\end{figure}

\subsubsection{Trapping region of the Chen system}
Obtaining a trapping region for the Lorenz system was quite simple, given the ellipsoidal Lyapunov function (without clear origins). The existence of the trapping region of the Chen system was proved in \cite{barbozachen}. The trapping region is the union of a series of ellipsoids.

The following proposition regarding the trapping region of the Chen system is taken from \cite{barbozachen}.
\begin{proposition}\label{thm:chenbounded}
    For any $Z_0 \leq 0$ and $U_0 \geq U_{0,\min}$ consider the ellipsoids
    \[E_i : \frac12 \left(\frac{A-Z_i}{A} X^2 + Y^2 + (Z-Z_i)^2 = U_i\right)\]
    $i=0,1,2,\dots$
where for $i \geq 1$
\[Z_i=Z_{i-1}\left(1+\frac{B}{AC} - \frac{B-1}{C}\right),\]
\[U_i = \frac{A-Z_i}{A-Z_{i-1}} \left[U_{i-1} - \frac{(Z_i-Z_{i-1})^2}{2}\right],\]
with $B>A/(A-Z_0)$. In the upper half-space $H_i:Z>Z_i$ let $\Omega_i=E^V_i\cap H_i$ where $E^V_i$ is the region embraced by $E_i$. Moreover for $N\geq 0$ define
\begin{equation}
    \Omega_{0,1,\dots,N}=\cup_{i=0}^{N} \Omega_i
\end{equation}
Then, there exists a non-negative integer $N$ such that $\Omega_{0,1,\dots,N}$ is a trapping region. The value of $N$ is determined by the condition
\begin{equation}
    U_N \leq \frac{1}{2}\left(\frac{B-1}{C}-Z_N\frac{B}{AC}\right)^2
\end{equation}
or, equivalently,
\begin{equation}
    \frac{B^2}{2C^2A^2}(A-Z_0)\left(A-\frac{A}{B} - Z_0\right)^2\sum_{i=0}^N\frac{\sigma^{2i}}{\frac{A}{B}+\left(A-\frac{A}{B} - Z_0\right)\sigma^i}
\end{equation}
where $\sigma=1+B/(AC)$.
\end{proposition}

Using proposition \ref{thm:chenbounded} with $Z_0 = 0$, $U_0 = U_{0, \min} = (D-Z_0)^2/2$, and the original parameters $a = 35$, $b = 3$, $c = 28$ one can obtain a trapping region as the union of 38 ellipsoids. The minimal $x, y, z$ coordinates of the trapping region are roughly $(-18.781, -48.1914, -271.508)$. We applied a shift of $(50, 50, 300)$ for easier to read equations and to maintain the generating reaction system at order 4. Choosing the initial concentrations in the ellipsoid $(x-50)^2 + (y-50)^2 + (z-300)^2 = 2 U_0 = 352.742$, the first ellipsoid, shifted, in the series of ellipsoids ensures that the trajectories start inside the trapping region.

\begin{figure}
    \centering
    \includegraphics[trim={0 4cm 0 1cm},clip,width=0.6\linewidth]{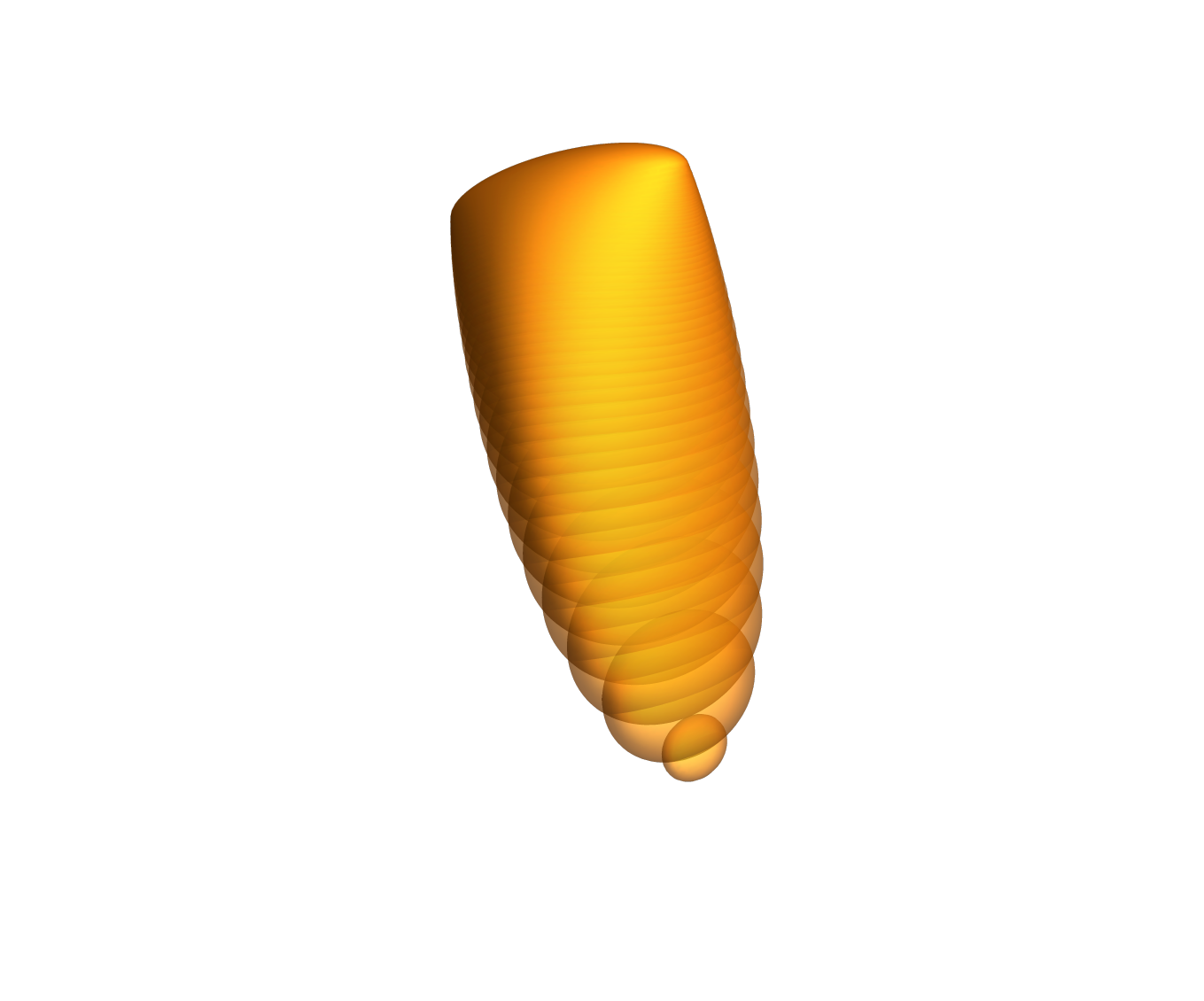}
    \caption{A trapping region of the Chen system with $Z_0 = 0$. Union of 38 ellipsoids.}
\end{figure}

\subsubsection{The transformed Chen system and its realization}
There is no need to give here the shifted equations and the corresponding figures, see the Electronic Supplement. Here we only present the canonic realization 
\begin{align}
&\ce{2 Y + Z ->[10] X + 2 Y + Z},\quad
\ce{X + Y + Z ->[10] Y + Z},\label{eq:chen01}\\
&\ce{X + Y + Z ->[328] X + 2 Y + Z},\quad
\ce{Y + 2 Z ->[50] 2 Y + 2 Z},\label{eq:chen02}\\
&\ce{X + Y + 2 Z ->[1] X + 2 Z},\quad
\ce{2 Y + Z ->[1] Y + Z},\label{eq:chen03}\\
&\ce{Y + Z ->[16350] Z},\quad
\ce{X + 2 Y + Z ->[1] X + 2 Y + 2 Z},\label{eq:chen04}\\
&\ce{Y + Z ->[3300] Y + 2 Z},\quad
\ce{X + Y + Z ->[50] X + Y},\label{eq:chen05}\\
&\ce{2 Y + Z ->[50] 2 Y},\quad
\ce{Y + 2 Z ->[8/3] Y + Z},\label{eq:chen06}
\end{align}
the shifted and multiplied equations: 
\begin{align}
X' &=  -10 X Y Z - 10 Y^2 Z \label{eq:ChenMultiplied01} \\
Y' &= -16350 Y Z + 328 X Y Z - Y^2 z + 50 Y Z^2 - X Y Z^2 \label{eq:ChenMultiplied02}\\
Z' &= 3300 Y Z - 50 X Y Z - 50 Y^2 Z + X Y^2 Z - 8/3 Y Z^2.\label{eq:ChenMultiplied03}
 \end{align}

and the corresponding figures in Fig. \ref{fig:ChenMultiplied}.
 \begin{figure}[!ht]
\centering
\includegraphics[width=0.45\linewidth]{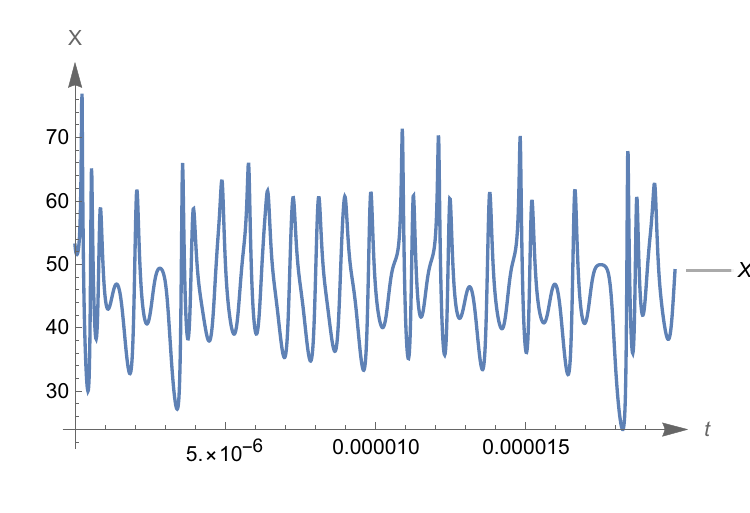}\quad
\includegraphics[width=0.45\linewidth]{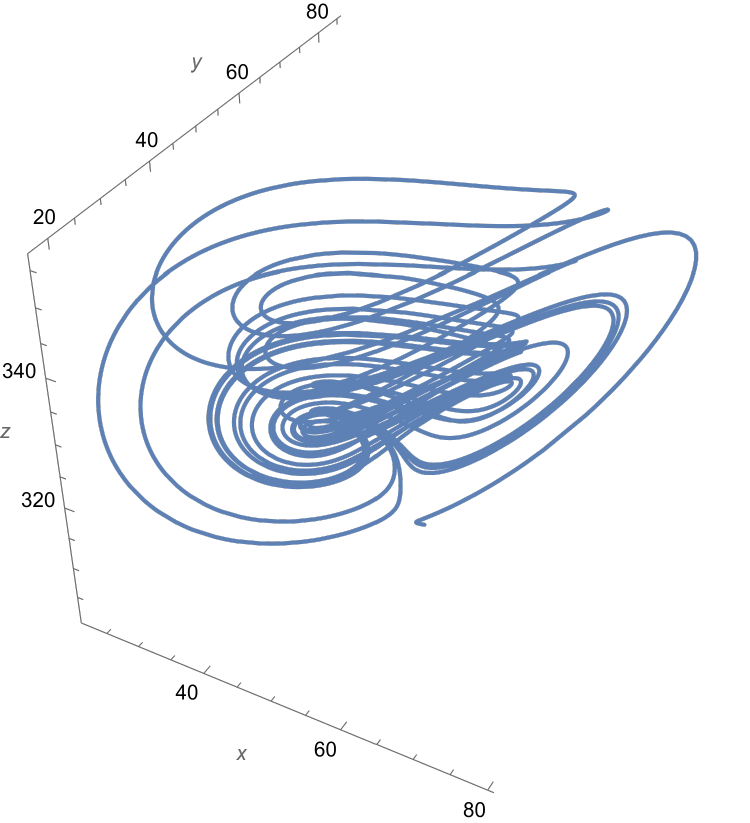}
\caption{Time evolution of the first component in the Chen reaction (left) and the corresponding trajectory (right) started from $(53,51,304)$ on the time interval 0-13.}
\label{fig:ChenMultiplied} 
\end{figure}
We also show the FHJ graph of the Chen reaction in Fig. \ref{fig:ChenFHJ}.
\begin{figure}[!ht]
\centering
\includegraphics[width=0.95\linewidth]{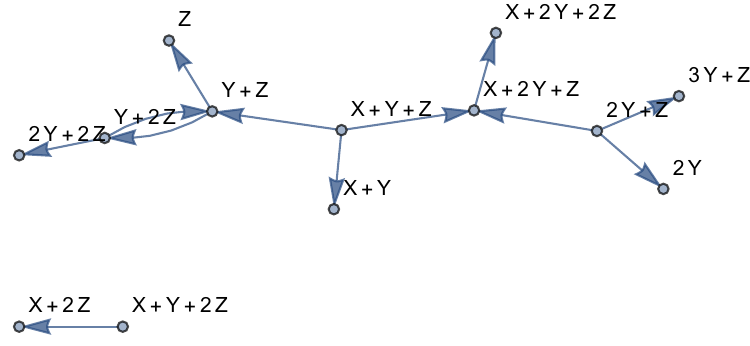}
\caption{The Feinberg--Horn--Jackson graph of the reaction \eqref{eq:chen01}--\eqref{eq:chen06}.
}
\label{fig:ChenFHJ} 
\end{figure}

\section{Discussion and outlook}\label{sec:disc}
The usual non-rigorous approach consists of finding that some properties seem to hold approximately. 
Given a differential equation, it is looked for whether the components of the solution or the trajectories fulfill some of the properties mentioned above or in the Appendix.
This has been done by \cite{rosslersimplereaction,rosslerabstract}, \cite{gaspard} \cite{poland} and \cite{wangxiao}.
Experimental results are evaluated similarly.

Here we are interested in a formal proof for induced kinetic differential equations of complex chemical reactions. 
As no such equation has rigorously shown chaotic behavior in the sense of the definitions, we did the following: we took for granted that the Lorenz model (it has a Smale horseshoe) and 
the Chen model (satisfies the Shil'nikov criterion, therefore it has a Smale horseshoe) are chaotic. Then, we were looking for reactions that have the same trajectories.

The difference between a proof based on induction as opposed to rigorous proof, 
is shown here by examples when the solutions may seem to show chaotic behavior on a certain finite interval only, but not on the whole real line.
Let us cite \cite{tel}: „Any system moving irregularly over a period of time and then changing to a regular behavior might be a candidate of transient chaos, cf. also  \cite{laitel,grebogiottyorke}.

Our method consists of similar steps to those applied earlier by other authors in heuristic investigations.
\begin{enumerate}[1]
\item
We shifted the trajectories into the first orthant.
Here the problem is to find the appropriate shift exactly.
\item
We multiplied the right-hand side of the equations with the product of some of the variables to eliminate negative cross-effect so as not to change the trajectories. This can be proven similarly to \cite[Section 3.1, Theorem 1]{perko}.
\item
We constructed a complex chemical reaction inducing the differential equation obtained in the previous step.
In other words, we found a realization of the obtained kinetic differential equation in terms of reaction steps endowed with mass action type kinetics.
\end{enumerate}
Our result here is a solution to a 50-year-old, never explicitly formulated, problem. However, researchers (\cite{willamowskirossler,poland} repeatedly tried to fill in the gap in the table below.
\cite{mendezgonzalesquezadatellezfernandezanayafemat} applied the Poland method of slow and fast reactions to approximately realize
a non-kinetic differential equation with reaction steps.

In all of our examples, we worked with the most often used \emp{standard parameter set}, it is only a long-range research project that could map the parameter space, 
as it can often be done when investigating multistationarity and periodicity.
We note that in some cases we utilized the possibility of free choice of the shifting parameters to reduce the order of the reaction steps in the given realization.

As to the realizations, first, note that this construction is by far not unique \cite{harstoth,craciunjohnstonszederkenyitonellotothyu}. 
Thus, one could either simplify the given reaction or find a simpler one. 
Therefore, we are left with an open problem of finding simpler, chemically more feasible realizations, those that are weakly reversible or have a low deficiency, etc. 
The works by \cite{szederkenyi,johnstonsiegelszederkenyi,deshpande} can help in this. 
Nevertheless, our work is a small contribution to the efforts looking for connections between the structure of complex chemical reactions and their dynamic behavior. 
As to having reaction steps of higher order, the situation is similar to the construction of a formal reaction with an arbitrary number of limit cycles \cite{erbankang}, or \cite{plesavejchodskyerban}: as a first step, it could only be achieved with less realistic reaction steps. 
Other chemical systems not described by mass action type kinetics may also show chaotic behavior, e.g. gas reactions \cite{davieshalfordmawhilltinsleyjohnsonscottkissgaspar} or
electrochemical reactions \cite{kissgasparnyikosparmanda,kissnagygaspar}.
Another related problem is to realize chaotic differential equations with electric circuits, see  \cite{cardellitribastonetschaikowski}.

One way to improve our approach is to make the chaotic equations chemically realistic. The order of the polynomial on the right-hand side should be at most 3. Better results can not be expected from directly applying our method as the starting equation has a second-degree right-hand side and the multiplier is degree 3 which leads to order 5 reaction steps. Choosing the shift well we could reduce this to order 4 reaction steps.  

An alternative approach to rigorously prove the existence of chaos in reactions is to start the investigation of a kinetic model from scratch and follow the steps described in e.g. \cite{zgliczynski,galiaszgliczynski}, as it happened with several non-kinetic (and non-kinetic, so far!) examples; surely not an easy way. 
A good idea in the paper \cite{ferreiraferreiralinoporto} is to check aperiodicity by looking at the Fourier transform of the concentrations and see that there are no determined peaks in it.

Our Mathematica/Wolfram Language notebooks can be found on \href{https://github.com/susitsm/ChaosInChemicalKinetics.git}{GitHub}.

\nocite{*}
\bibliographystyle{apalike}
\bibliography{_ChemicalChaos}
\newpage
\section{Appendix}\label{sec:appendix}
In the Appendix, we provide a small review of formal reaction kinetics and chaos. They are not a substitute for a textbook. 
\subsection{Induced kinetic differential equations}\label{subsec:ikde}
We are going to use the following concepts of \emp{formal reaction kinetics}, also called \emp{chemical reaction network theory}.
Following the books by \cite{feinbergbook} and by \cite{tothnagypapp} we consider a \emp{complex chemical reaction}, simply \emp{reaction}, or \emp{reaction network} as a set consisting of \emp{reaction steps} as follows: 
\begin{equation}\label{eq:ccr}
\left\{\ce{$\sum_{m=1}^M\alpha_{m,r}$X($m$) -> $\sum_{m=1}^M\beta_{m,r}$X($m$)};\quad (r=1,2,\dots,R)\right\}; 
\end{equation}
where 
\begin{enumerate}
\item 
the chemical \emp{species} are \ce{X($1$)}, \ce{X($2$)}, \dots, \ce{X($M$)};
\item
the \emp{reaction steps} are numbered from 1 to \(R;\) 
\item
here \(M\) and \(R\) are positive integers;
\item
\(\alphab:=[\alpha_{m,r}]\) and 
\(\betab:=[\beta_{m,r}]\) are \(M\times R\) matrices of non-negative integer components called \emp{stoichiometric coefficients}, with the properties that all the species take part in at least one reaction step (\(\forall m \exists r: \beta_{m,r}\neq\alpha_{m,r}\)), and all the reaction steps do have some effect (\(\forall r \exists m: \beta_{m,r}\neq\alpha_{m,r}\)), and finally
\item
\(\gammab:=\betab-\alphab\) is the \emp{stoichiometric matrix} of  \emp{stoichiometric numbers}.
\end{enumerate}
We now provide a simple example to make the understanding easier. 
\begin{equation}
\ce{Y <=> 0},\quad\ce{0 <=> X + Z},\quad\ce{2 Y <=> X + Y},\quad\ce{2Z <=>  Z},\quad
\ce{2X <=>  X}.\label{eq:willross}    
\end{equation}
In Eq. \eqref{eq:willross} taken from \cite{willamowskirossler} by neglecting the external components one has \(M=3\) species and \(R=10\) reaction steps.
The matrices of stoichiometric coefficients are
\begin{equation*}
\alphab=
\begin{bmatrix}
0&0&0&1&0&1&0&0&2&1\\
1&0&0&0&2&1&0&0&0&0\\
0&0&0&1&0&0&2&1&0&0
\end{bmatrix},\quad
\betab=\begin{bmatrix}
0&0&1&0&1&0&0&0&1&2\\
0&1&0&0&1&2&0&0&0&0\\
0&0&1&0&0&0&1&2&0&0
\end{bmatrix}.
\end{equation*}
\begin{figure}[!ht]
\centering
\includegraphics[width=0.85\linewidth]{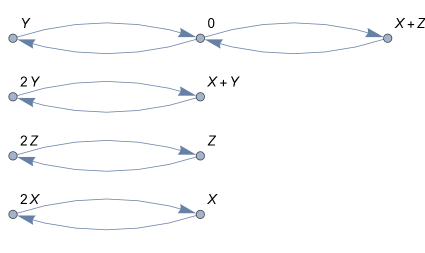}
\caption{The Feinberg--Horn--Jackson graph of the complex chemical reaction \eqref{eq:willross}.
}
\label{fig:willrossfhj} 
\end{figure}
The number \(N\) of formal linear combinations on both sides of the reaction arrows, called \emp{complexes} is 9, 
the number of connected components of the graph 
(the \emp{Feinberg--Horn--Jackson graph}, see Fig. \ref{fig:willrossfhj}) with vertices as complexes and with edges as reaction arrows is 4,
finally, the rank \(S\) of the stoichiometric matrix is 3. 
The deficiency \(\delta\) of this complex chemical reaction is \(\delta:=N-L-S=9-4-3=2.\)

The time evolution of the concentrations of the chemical species will be described by the \emp{induced kinetic differential equation} 
\begin{equation}\label{eq:ikdegen}
\dot{\cb}(t)=\gammab\rateb(\cb(t))
\end{equation} 
together with the initial condition \(\cb(0)=\cb^0.\)

The component \(rate_r\) of the vector \(\rateb\) provides the reaction rate of the \(\rth\) reaction step. We assume here the
\emp{mass-action} case, then Eq. \eqref{eq:ikdegen} specializes into
\begin{equation}\label{eq:ikdemass}
\dot{\cb}=\gammab\kb\odot\cb^{\alphab} 
\end{equation} 
or, in coordinates
\begin{equation*}
\dot{c}_m(t)=\sum_{r=1}^{R}\gamma_{mr}k_r\prod_{p=1}^{M}c_p^{\alpha_{p,r}}\quad(m=1,2,\cdots,M),
\end{equation*}
where \(\kb\) is the vector of (positive) reaction rate coefficients \(k_r.\)
In Eq. \eqref{eq:ikdemass} we used the usual vector operations, see e.g. Sec. 13.2 of \cite{tothnagypapp}.
Their use in formal reaction kinetics has been initiated by \cite{hornjackson}.
Eq. \eqref{eq:ikdemass} is a special polynomial differential equation. Equations arising as the induced kinetic differential equations can fully be characterized within the class of polynomial differential equations via the Hungarian lemma \cite{harstoth}.
\begin{lemma}\label{lemma:inverse}
The polynomial differential equation
\begin{equation}
\dot{c}_m=p_m(\cb(t))\quad (m=1,2,\dot,M)    
\end{equation}
can be realized by a complex chemical reaction (it is \emp{kinetic} or \emp{Hungarian}) if and only if
\begin{equation}
p_m(\cb)=f_m(\cb)-c_mg_m(\cb)\quad (m=1,2,\dots,M)    
\end{equation}
holds where \(f_m\) and \(g_m\) are polynomials with nonnegative coefficients.
\end{lemma}
In plain words,  Lemma \ref{lemma:inverse} expresses the fact that the concentration of no species can decrease in a process in which it does not take part. If one has a kinetic differential equation then one can automatically construct one realization of the equation in terms of formal reactions. This realization is not simple, not optimal in any sense, further refining may give nicer ones.
\begin{definition}[Canonic realization]
The canonic realization is as follows.
A kinetic differential equation has two types of terms on its \(m^{th}\) right-hand side.
\begin{enumerate}
\item[\nonumber]
\item 
Terms of the form \(k\prod_{p=1}^Mc_p^{\alpha_p}, \quad k>0\) can be realized by the reaction step 
\begin{equation}
\ce{$\sum_{p=1}^M\alpha_p$ X($p$) ->[$k$] $\sum_{p=1}^M(\alpha_p+\delta_{m,p})$ X($p$)}.
\end{equation}
\item 
Terms of the form \(-k\prod_{p=1}^Mc_p^{\alpha_p}, \quad k>0\) can be realized by the reaction step 
\begin{equation}\label{eq:secondreal}
\ce{$\sum_{p=1}^M\alpha_p$ X($p$) ->[$k$] $\sum_{p=1}^M(\alpha_p-\delta_{m,p})$ X($p$)}.
\end{equation}
\end{enumerate}
\end{definition}
The stoichiometric coefficients of the product complex are non-negative because the differential equation is kinetic.
\subsection{Characteristics of chaotic behavior}\label{subsec:chaos}
As no universally accepted definition of chaos exists, we mention a few possible approaches following \cite[Subsection 1.2.5]{bishop}. 
We do not even mention the (far from simple) relationships between the different definitions.
\begin{definition}[Devaney]\label{def:devaney}
A continuous map \(f\) is chaotic if \(f\) has an invariant subset \(K\) of the state space such that
\begin{enumerate}
\item	
\(f\) satisfies weak sensitive dependence on \(K\),
\item	
The set of points initiating periodic orbits are dense in \(K\), and
\item	
\(f\) is topologically transitive on \(K\).
\end{enumerate}
\end{definition}
\begin{definition}\label{def:toptrans}
Topological transitivity means that no matter how small open sets \(U\) and \(V\) are, some trajectory starting from \(U\) eventually visits \(V\).
\end{definition}
\begin{definition}[Smale, horseshoe]\label{def:smale}
A discrete map is chaotic if after some iteration it maps the unit interval into a horseshoe.
\end{definition}
This implies Definition \ref{def:devaney}. 
It implies sensitive dependence.
\begin{definition}[Smith, topological entropy]
The topological entropy is positive.
\end{definition}
Let \(f\) be a discrete map and \(\{W_i\}\)
be a partition of a bounded region \(W\)
containing a probability measure that is invariant under \(f\).
\begin{definition}\label{topentr}
The \emp{topological entropy} of \(f\)
is defined as \(h_T(f):=\sup_{W_i}h(f,\{W_i\}),\)
where \(\sup\) is the supremum on the set \(\{Wi\}.\)
\end{definition}
\begin{definition}[Physics, Lyapunov]\label{def:physicslyapunov}
A discrete map is chaotic if it has a positive global Lyapunov exponent.
\end{definition}
\begin{definition}[Period doubling]\label{def:perioddoubling}
A period-doubling bifurcation occurs when a slight change in a system's parameters causes a new periodic trajectory to emerge from an existing trajectory---the new one having double the period of the original. With the double period, it takes twice as long (or, in a discrete dynamical system, twice as many iterations) for the numerical values visited by the system to repeat itself. 
\end{definition}
\end{document}